\documentclass[]{article}

\usepackage{graphicx}
\usepackage[colorlinks=true,breaklinks=true,citecolor=blue,linkcolor=blue,urlcolor=blue]{hyperref}
\usepackage{lscape}
\usepackage[german, english]{babel}
\usepackage{amssymb}
\usepackage{amsmath}
\usepackage[top=2cm,left=2cm,right=2cm,bottom=2cm]{geometry} 

\title{\textsl{\textbf{Energy and matter}}}

\author{\textbf{Ricardo Gobato$^{1}$}\\ Secretaria de Estado da Educa\c{c}\~ao do Paran\'{a} (SEED/PR),\\ Av. Maring\'{a}, 290, Jardim Dom Bosco,\\ Londrina/PR, 86060-000, Brasil\\ \\
	\textbf{Alekssander Gobato$^{2}$}\\ Faculdade Pit\'{a}goras Londrina, \\Rua Edwy Taques de Ara\'{u}jo, 1100,\\ Gleba Palhano, Londrina/PR, 86047-500, Brasil\\ \\
	\textbf{Desire Francine Gobato Fedrigo$^{3}$}\\  Aeronautical Engineering Consulting\\ Consultant in processes LOA/PBN RNAV, Rua Lu\'{i}sa, 388s, ap. 05,\\ Vila Portuguesa, Tangar\'{a} da Serra/MT, 78300-000, Brasil\\ \\
	\textbf{Corresponding authors}:\\ $^{1}$ricardogobato@seed.pr.gov.br;\\ $^{2}$alekssandergobato@hotmail.com;\\
	$^{3}$desirefg@bol.com.br}

\begin{document}

\maketitle
 
\begin{center}
	\textbf{Keywords}: {Acceleration of gravity, Einstein, \textit{E} = \textit{mc}$^{2}$, Energy, Matter, Universal Gravitation.}
\end{center}

\begin{abstract}
Our work is an approach between matter and energy. Using the famous equation \textit{E} = \textit{mc}$^{2}$, Einstein and the Law of Universal Gravitation of Newton, we estimate that a small amount matter converted into energy is needed to lift, using the gravitational potential energy equation on the surface, a mountain of solid iron or even Mount Everest.
 \end{abstract}

\twocolumn[]
\section{Introduction}

We work in interpreting a connection between energy-matter, based on the parable and the famous equation \textit{E} = \textit{mc}$^{2}$, which by this equation matter and energy are already connected. Whereas a small amount of material, such as 10 mg mass, considering that all material is converted to energy. How high could lift a mountain of solid iron, or even Mount Everest, with this energy? Energy is energy, no matter what form it is, it is used to produce work, or move something.
Whereas the great mountains like Mount Everest, analyzing altitude, size, density, etc. We used the iron, to do a comparison of the energy given by Einstein's equation and gravitational potential energy. We approach the law of universal gravitation and the speed of light.\\

We made a comparison of the energy needed to lift a massive iron mountain and Mount Everest. A mountain of solid iron to get an idea of its huge mass. Mount Everest because it is the highest mountain on earth.

\section{Mount Everest}
 
Mount Everest, also known in Nepal as Sagarmāthā and in Tibet as Chomolungma, is Earth's highest mountain. It is located in the Mahalangur section of the Himalayas. Its peak is 8,848 metres (29,029 ft) above sea level \cite{2010}. It is not the furthest summit from the centre of the Earth. That honour goes to Mount Chimborazo, in the Andes \cite{Krulwich2007}. The international border between China and Nepal runs across Everest's precise summit point. Its massif includes neighbouring peaks Lhotse, 8,516 m (27,940 ft); Nuptse, 7,855 m (25,771 ft) and Changtse, 7,580 m (24,870 ft).\

In 1856, the Great Trigonometrical Survey of India established the first published height of Everest, then known as Peak XV, at 29,002 ft (8,840 m). The current official height of 8,848 m (29,029 ft) as recognised by China and Nepal was established by a 1955 Indian survey and subsequently confirmed by a Chinese survey in 1975. In 1865, Everest was given its official English name by the Royal Geographical Society upon a recommendation by Andrew Waugh, the British Surveyor General of India. Waugh named the mountain after his predecessor in the post, Sir George Everest, arguing that there were many local names, against the opinion of Everest. \cite{1857}\

Mount Everest attracts many highly experienced mountaineers as well as capable climbers willing to hire professional guides. There are two main climbing routes, one approaching the summit from the southeast in Nepal (known as the standard route) and the other from the north in Tibet. While not posing substantial technical climbing challenges on the standard route, Everest presents dangers such as altitude sickness, weather, wind as well as significant objective hazards from avalanches and the Khumbu Icefall.\

The first recorded efforts to reach Everest's summit were made by British mountaineers. With Nepal not allowing foreigners into the country at the time, the British made several attempts on the north ridge route from the Tibetan side. After the first reconnaissance expedition by the British in 1921 reached 7,000 m (22,970 ft) on the North Col, the 1922 expedition pushed the North ridge route up to 8,320 m (27,300 ft) marking the first time a human had climbed above 8,000 m (26,247 ft). Tragedy struck on the descent from the North col when seven porters were killed in an avalanche. The 1924 expedition resulted in the greatest mystery on Everest to this day: George Mallory and Andrew Irvine made a final summit attempt on June 8 but never returned, sparking debate as to whether they were the first to reach the top. They had been spotted high on the mountain that day but disappeared in the clouds, never to be seen again until Mallory's body was found in 1999 at 8,155 m (26,755 ft) on the North face. Tenzing Norgay and Edmund Hillary made the first official ascent of Everest in 1953 using the southeast ridge route. Tenzing had reached 8,595 m (28,199 ft) the previous year as a member of the 1952 Swiss expedition. \cite{Rogers, Mount-everest}\

\begin{figure}
	\begin{center}
		\includegraphics[scale=0.8]{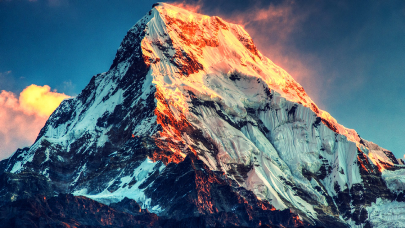}
		\caption{\small{Mount Everest, also known in Nepal as \textit{Sagarm\-{a}th\-{a}} and in Tibet as Chomolungma, is Earth's highest mountain. It is located in the Mahalangur section of the Himalayas. Its peak is 8,848 metres (29,029 ft) above sea level \cite{everest1}.}}\label{fig:sweeps}
	\end{center}
\end{figure}

\subsubsection{Expeditions to Mount Everest}

Everest Base Camp is a term that is used to describe two base camps on opposite sides of Mount Everest. South Base \ Camp is in Nepal at an altitude of 5,364 metres (17,598 ft) (28$^{\circ}$0'26"N 86$^{\circ}$51'34"E), and North Base Camp is in Tibet at 5,150 metres (16,900 ft) \cite{Wing-Sze2010, Reynolds2006, Buckley2008} (28$^{\circ}$8'29"N  86$^{\circ}$51'5"E). These camps are rudimentary campsites on Mount Everest that are used by mountain climbers during their ascent and descent. South Base Camp is used when climbing via the southeast ridge, while North Base Camp is used when climbing via the northeast ridge. \cite{Mayhew2009, basecampeverest}\\

Supplies are carried to the South Base Camp by sherpas or porters, and with the help of animals, usually yaks. The North Base Camp has vehicle access (at least in the summer months). Climbers typically rest at base camp for several days for acclimatization to reduce the risks and severity of altitude sickness.\\

\begin{figure}
	\begin{center}
		\includegraphics[scale=0.66]{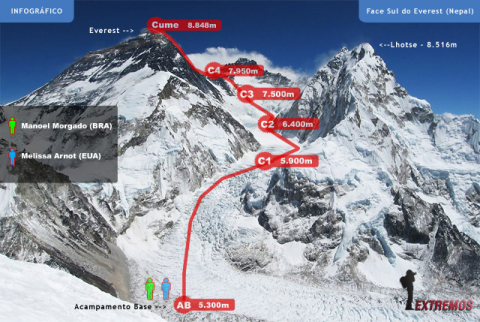}
		\caption{\small{Layout of the mount everest base camp, traveled by mountaineer Rodrigo Raineri base camp to advanced Field 2 the highest mountain in the world - located 6,400 meters high. Base camp located at 5,300 meters. \cite{Cesar}}}\label{fig:sweeps}
	\end{center}
\end{figure}

\textbf{South Base Camp in Nepal}\\

The Everest Base Camp trek on the south side is one of the most popular trekking routes in the Himalayas and is visited by thousands of trekkers each year. Trekkers usually fly from Kathmandu to Lukla to save time and energy before beginning the morning trek to this base camp. However, trekking to Lukla is possible.By 2015 is was noted about 40 thousand people year take the trek from Lukla airport to the Nepal Everest Base Camp \cite{Foxnews2015}. From Lukla, climbers trek upward to the Sherpa capital of Namche Bazaar, 3,440 metres (11,290 ft), following the valley of the Dudh Kosi river. It takes about two days to reach the village, which is a central hub of the area. Typically at this point, climbers allow a day of rest for acclimatization. They then trek another two days to Dingboche, 4,260 metres (13,980 ft) before resting for another day for further acclimatization. Another two days takes them to Everest Base Camp via Gorakshep, the flat field below Kala Patthar, 5,545 metres (18,192 ft) and Mt. Pumori. \cite{basecampeverest}\\

On 25 April 2015 an earthquake measuring 7.8 Mw struck Nepal and triggered an avalanche on Pumori that swept through the South Base Camp \cite{NationalGeographic}. At least 19 people were said to have been killed as a result. Just over two weeks later, on May 12, a second quake struck measuring 7.3 on the Richter scale. \cite{timesindia}\\

In 2015, some of these trails were damaged by the earthquake and needed to be repaired. \cite{Foxnews2015}\\

\section{Iron}

Iron is a chemical element with symbol Fe (from Latin: ferrum) and atomic number 26. It is a metal in the first transition series \cite{Pure2013}. It is by mass the most common element on Earth, forming much of Earth's outer and inner core. It is the fourth most common element in the Earth's crust. Its abundance in rocky planets like Earth is due to its abundant production by fusion in high-mass stars, where the production of nickel-56 (which decays to the most common isotope of iron) is the last nuclear fusion reaction that is exothermic. Consequently, radioactive nickel is the last element to be produced before the violent collapse of a supernova scatters precursor radionuclide of iron into space.\

\begin{figure}
	\begin{center}
		\includegraphics[scale=1.185]{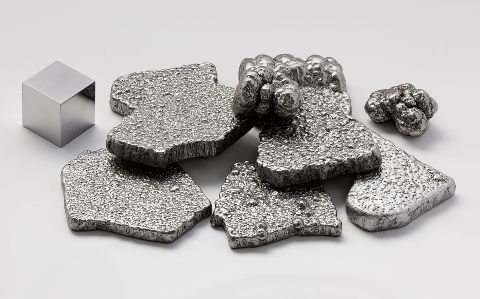}
		\includegraphics[scale=0.244]{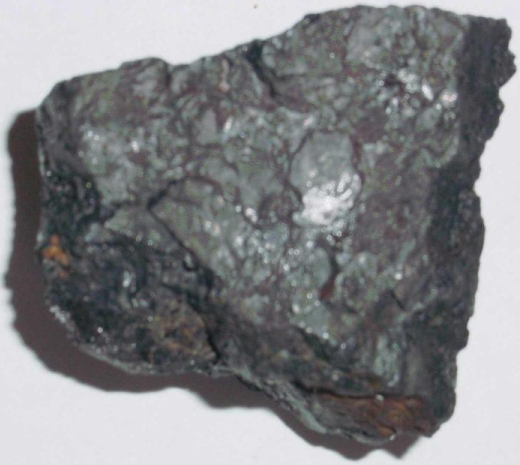}\\
		\includegraphics[scale=0.4]{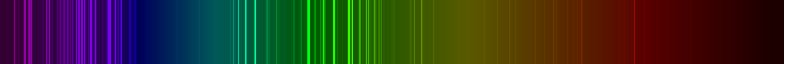}
		\caption{\small{Iron electrolytic and 1cm$^{3}$ cube. Iron ore. Ironstone is a fine-grained, heavy and compact sedimentary rock. Its main components are the carbonate or oxide of iron, clay and/or sand. It can be thought of as a concretionary form of siderite. Ironstone also contains clay, and sometimes calcite and quartz. \cite{iron, ironore, IUPAC}}}\label{fig:sweeps}
	\end{center}
\end{figure}

Like other group 8 elements, iron exists in a wide range of oxidation states, −2 to +6, although +2 and +3 are the most common. Elemental iron occurs in meteoroids and other low oxygen environments, but is reactive to oxygen and water. Fresh iron surfaces appear lustrous silvery-gray, but oxidize in normal air to give hydrated iron oxides, commonly known as rust. Unlike many other metals which form passivating oxide layers, iron oxides occupy more volume than the metal and thus flake off, exposing fresh surfaces for corrosion.\

\subsection{Physico-chemical properties of iron}

Atomic number: 26; Standard atomic weight ($\pm$) (Ar) 55.845(2) \cite{IUPAC}; Element category: transition metal; Group, block: group 8, d-block; Period: period 4; Electron configuration:
[Ar] 3d$^{6}$ 4s$^{2}$; per shell	2, 8, 14, 2\\

\textbf{Physical properties}\\

Phase: solid; Melting point: 1811 K (1538 $^{\circ}$C, 2800 $^{\circ}$F); Boiling point
3134 K (2862 $^{\circ}$C, 5182 $^{\circ}$F); Density near: 7.874 g/cm$^{3}$; when liquid, at m.p.: 6.98 g/cm$^{3}$; Heat of fusion: 13.81 kJ/mol; Heat of vaporization: 340 kJ/mol; Molar heat capacity: 25.10 J/(mol.K)\\

\textbf{Atomic properties}\\

Oxidation states: −2, −1, +1, \cite{Ram2003} +2, +3, +4, +5, \cite{Pouchard1982} +6 (an amphoteric oxide)\\

\textbf{Electronegativity}\\

Pauling scale: 1.83

\textbf{Ionization energies}\\
1st: 762.5 kJ/mol; 2nd: 1561.9 kJ/mol; 3rd: 2957 kJ/mol\\

Atomic radius: empirical: 126 pm\

Covalent radius: Low spin: 132±3 pm; High spin: 152±6 pm\\

\textbf{Crystal structure}\\

Body-centered cubic (bcc): a=286.65 pm; face-centered cubic (fcc):
between 1185 - 1667 K\

Speed of sound thin rod: 5,120 m/s (at r.t.) (electrolytic);
Thermal expansion: 11.8 µm/(m.K) (at 25 $^{\circ}$C); 
Thermal conductivity: 80.4 W/(m.K); 
Electrical resistivity: 96.1 nΩ.m (at 20 $^{\circ}$C); 
Curie point: 1043 K; Magnetic ordering: ferromagnetic; Young's modulus: 211 GPa; Shear modulus: 82 GPa; Bulk modulus: 170 GPa; Poisson ratio: 0.29; Mohs hardness: 4; Vickers hardness: 608 MPa;
 Brinell hardness: 200 - 1180 MPa; CAS Registry Number: 7439-89-6; 
History (Discovery): before 5000 BC\

\section{Fundamentals of Physical}

\subsection{Potential Energy}

The total work done on a particle equals the change in its kinetic energy. But we are frequently interested in the work done on a system of two or more particles. Often, the work done by external forces on a system does not increase the kinetic energy of the system, but instead is stored as potential energy. \cite{Tipler1999, Resnick}\

\subsubsection{Conservative Forces}

When you ride a ski lift to the top of a hill of height h, the work done bythe lift on you is mgh and that done by gravity is \textit{–mgh}. When you ski down the hill to the bottom, the work done by gravity is \textit{+mgh}
independent of the shape of the hill. The total work done by gravity on you during the round trip is zero independent of the path you take. The force of gravity exerted by the earth on you is called a \textit{conservative force}.\\

\textbf{Definition—Conservative force}\\

A force is conservative if the total work it does on a particle is zero when the particle moves around any closed path returning to its initial position.\\

\textbf{Alternative definition—Conservative force}\\

The work done by a conservative force on a particle is independent of the path taken as the particle moves from one point
to another. \cite{Tipler1999, Resnick}\

\subsubsection{Potential-Energy Functions}

Since the work done by a conservative force on a particle does not depend on the path, it can depend only on the endpoints 1 and 2. We can use this property to define the \textit{potential-energy function U} that is associated with a conservative force. Note that when the skier skis down the hill, the work done by gravity \textit{decreases} the potential energy of the system. In general, we define the potential energy function such that the work done by a conservative force equals the decrease in the potentialenergy function:

\begin{equation}
\
W=\int {\bar{F}\, .\vec{ds}} =\, -\Delta U
\
\end{equation}

or Definition—Potential-energy function, Equation (2)

\begin{equation}
\
\Delta U=U_{2}-U_{1} = -\int^{S_2}_{S_1} {\bar{F}\, .\vec{ds}} 
\
\end{equation}

For infinitesimal displacement, we have

\begin{equation}
\
dU=\, -\bar{F}\, .\vec{ds}
\
\end{equation}

\subsubsection{Gravitational Potential Energy Near the Earth’s Surface}

We can calculate the potential-energy function associated with the gravitational force near the surface of the earth from Equation (3). For the force

\begin{equation}
\
\bar{F}=-mg\hat{j}
\
\end{equation}

we have

\[
dU=-\bar{F}.\, d\vec{s}=-\left( -m.g.\hat{j} \right).\, \left( 
dx\hat{i}+dy\hat{j}+dz\hat{k} \right)
\]
\begin{equation}
\
=+mg\, dy
\
\end{equation}

Integrating, we obtain Gravitational potential energy near the earth’s surface Equation (7)

\begin{equation}
\
U=\int {mg\, dy} =mgy+U_{o}
\
\end{equation}

\begin{equation}
\
U=U_{o}+mgy
\
\end{equation}

where \textbf{U}$_{0}$, the arbitrary constant of integration, is the value of the
potential energy at y = 0. Since only a change in the potential energy is defined, the actual value of \textit{U} is not important. We are free to choose \textit{U} to be zero at any convenient reference point. For example, if the gravitational potential energy of the earth–skier system is chosen to be zero when the skier is at the bottom of the hill, its value when the skier is at a height \textit{h} above that level is \textit{mgh}. Or we could choose the potential energy to be zero when the skier is at sea level, in which case its value at any other point would be \textit{mgy}, where y is measured from sea level. \cite{Tipler1999, Resnick}

\subsection{Newton’s Law of Gravity}

Although Kepler’s laws were an important first step in understanding the motion of planets, they were still just empirical rules obtained from the astronomical observations of Brahe. It remained for Newton to take the next giant step by attributing the acceleration of a planet in its orbit to a specific force exerted on it by the sun. Newton proved that a force that varies inversely with the square of the distance between the sun and a planet results in elliptical orbits, as observed by Kepler. He then made the bold assumption that this force acts between any two objects in the universe. Before Newton, it was not even generally accepted that the laws of physics observed on earth were applicable to the heavenly bodies. Newton’s law of gravity postulates that there is a force ofattraction between each pair of objects that is proportional to the product of the masses of the objects and inversely proportional to the square of
the distance separating them. The magnitude of the gravitational force exerted by a particle of mass \textit{m}$_{1}$ on another particle of mass \textit{m}$_{2}$ a distance \textit{r} away is thus given by

\begin{equation}
	\
	F=\, \frac{G.m_{1}m_{2}}{r^{2}}
	\
\end{equation}

where \textit{G} is the universal gravitational constant, which has the value

\begin{equation}
	\
	G=6.67 \times {10}^{-11}N{kg}^{2}/m^{2}
	\
\end{equation}

Newton published his theory of gravitation in 1686, but it was not until a century later that an accurate experimental determination of \textit{G} was made by Cavendish, whose findings will be discussed in the next section. If \textit{m}$_{1}$ is at position $\vec{r}$$_{1}$ and \textit{m}$_{2}$ is at $\vec{r}$$_{2}$, the force exerted by mass  \textit{m}$_{1}$ on \textit{m}$_{2}$ is

\begin{equation}
	\
	\bar{F}_{1,2}=\, \frac{G.m_{1}m_{2}}{r_{1,2}^{1}}\, \hat{r}_{1,2}
	\
\end{equation}

where $\vec{r}$$_{1,2}$ is the vector pointing from mass \textit{m}$_{1}$ to \textit{m}$_{2}$ and $\bar{r}$$_{1,2}$=$\vec{r}$$_{1,2}$/r$_{1,2}$ is
a unit vector point from \textit{m}$_{1}$ to \textit{m}$_{2}$. The force exerted by \textit{m}$_{2}$ on \textit{m}$_{1}$ is
the negative of , according to Newton’s third law.
We can use the known value of \textit{G} to compute the gravitational
attraction between two ordinary objects.

\cite{Tipler1999, Resnick}

\subsection{Gravitational and Inertial Mass}

The property of an object responsible for the gravitational force it exerts on another object is its
\textit{gravitational} mass, whereas the property of an object that measures its resistance to acceleration is
its \textit{inertial mass}. We have used the same symbol m for these two properties because, experimentally,
they are equal. The fact that the gravitational force exerted on an object is proportional to its inertial
mass is a characteristic unique to the force of gravity. One consequence is that all objects near the
surface of the earth fall with the same acceleration if air resistance is neglected. This fact has seemed
surprising to all since it was discovered. The famous story of Galileo demonstrating it by dropping
objects from the Tower of Pisa is just one example of the excitement this discovery aroused in the
sixteenth century.\\
We could easily imagine that the gravitational and inertial masses of an object were not the
same. Suppose we write \textit{m}$_{G}$ for the gravitational mass and \textit{m} for the inertial mass. The force exerted
by the earth on an object near its surface would then be

\begin{equation}
	\
	F=\, \frac{G.M_{E}m_{G}}{R_{E}^{2}}
	\
\end{equation}

where \textit{M}$_{E}$ is the gravitational mass of the earth. The free-fall acceleration of the object near the
earth’s surface would then be

\begin{equation}
	\
	a=\frac{F}{m}=\, \left( \frac{G.M_{E}}{R_{E}^{2}} \right)\frac{m_{G}}{m}
	\
\end{equation}

If gravity were just another property of matter, like color or hardness, it might be reasonable to
expect that the ratio \textit{m}$_{G}$/\textit{m} would depend on such things as the chemical composition of the object or
its temperature. The free-fall acceleration would then be different for different objects. The
experimental fact, however, is that a is the same for all objects. Thus, we need not maintain the
distinction between \textit{m}$_{G}$ and \textit{m} and can set \textit{m}$_{G}$ = \textit{m}. We must keep in mind, however, that the equivalence of gravitational and inertial mass is an experimental law, one that is limited by the
accuracy of experiment. Experiments testing this equivalence were carried out by Simon Stevin in
the 1580s. Galileo publicized the law widely, and his contemporaries made considerable
improvements in the experimental accuracy with which the law was established.\\
The most precise early comparisons of gravitational and inertial mass were made by Newton.
From experiments using simple pendulums rather than falling bodies, Newton was able to establish
the equivalence between gravitational and inertial mass to an accuracy of about 1 part in 1000.
Experiments comparing gravitational and inertial mass have improved steadily over the years. Their
equivalence is now established to about 1 part in 10$^{12}$. The equivalence of gravitational and inertial
mass is therefore one of the most well established of all physical laws. It is the basis for the principle
of equivalence, which is the foundation of Einstein’s general theory of relativity.
\cite{Tipler1999, Resnick}

\subsection{Gravitational Potential Energy}

Near the surface of the earth, the gravitational force exerted by the earth
on an object is constant because the distance to the center of the earth \textit{r} = \textit{R}$_{E}$ + \textit{h} is always approximately \textit{R}$_{E}$ for \textit{h} $\ll$ \textit{R}$_{E}$ . The potential energy of
an object near the earth’s surface is \textit{mg}(\textit{r} $-$ \textit{R}$_{E}$) = \textit{mgh}, where we have
chosen \textit{U = 0} at the earth’s surface, \textit{r = R}$_{E}$ . When we are far from the
surface of the earth, we must take into account the fact that the gravitational force exerted by the earth is not uniform but decreases as 1/r$^{2}$. The general definition of potential energy  gives 

\begin{equation}
	\
	dU=\, -\bar{F}\, .\vec{ds}
	\
\end{equation}

where is the force on a particle and is a general displacement of the particle. For the radial gravitational force given by we have

\[
dU=\, -\bar{F}\, .\vec{ds}=\, -F_{r}dr=-\left( -\frac{GM_{E}m}{r^{2}} 	\right)dr
\]

\begin{equation}
\
=+\frac{GM_{E}m}{r^{2}}dr
\
\end{equation}

Integrating both sides of this equation we obtain

\begin{equation}
	\
	U=-\frac{GM_{E}m}{r}+U_{0}
	\
\end{equation}

where \textit{U}$_{0}$ is a constant of integration. Since only changes in potential
energy are important, we can choose the potential energy to be zero at
any position. The earth’s surface is a good choice for many everyday
problems, but it is not always a convenient choice. For example, when
considering the potential energy associated with a planet and the sun,
there is no reason to want the potential energy to be zero at the surface of
the sun. In fact, it is nearly always more convenient to choose the
gravitational potential energy of a two-object system to be zero when the
separation of the objects is infinite. Thus, \textit{U}$_{0}$ is often a convenient
choice. Then Gravitational potential energy with \textit{U = 0} at infinite separation is

\begin{equation}
	\
	U\left( r \right)=-\frac{GMm}{r},\, \, \, \, \, \, \, U=0\, \, \, at\, \, \, 
	r=\, \infty 
	\
\end{equation}

\cite{Tipler1999, Resnick}

\subsection{Gravitational acceleration}

In physics, gravitational acceleration is the acceleration on an object caused by force of gravitation. Neglecting friction such as air resistance, all small bodies accelerate in a gravitational field at the same rate relative to the center of mass. This equality is true regardless of the masses or compositions of the bodies.\\

At different points on Earth, objects fall with an acceleration between 9.78 and 9.83 m/s$^{2}$ depending on altitude and latitude, with a conventional standard value of exactly 9.80665 m/s$^{2}$ (approximately 32.174 ft/s$^{2}$). Objects with low densities do not accelerate as rapidly due to buoyancy and air resistance.

 The force acting upon the smaller mass can be calculated as:
 
\begin{equation}
\
\vec{F} = m . \vec{g}
\
\end{equation}
 
where $\vec{F}$ is the force vector, \textit{m} is the smaller mass, and $\vec{g}$ is a vector pointed toward the larger body. Note that $\vec{g}$ has units of acceleration and is a vector function of location relative to the large body, independent of the magnitude (or even the presence) of the smaller mass.\\
 
This model represents the ``far-field" gravitational acceleration associated with a massive body. When the dimensions of a body are not trivial compared to the distances of interest, the principle of superposition can be used for differential masses for an assumed density distribution throughout the body in order to get a more detailed model of the "near-field" gravitational acceleration. For satellites in orbit, the far-field model is sufficient for rough calculations of altitude versus period, but not for precision estimation of future location after multiple orbits. \cite{Gravitational}\\

\subsection{Gravity model for Earth}

The type of gravity model used for the Earth depends upon the degree of fidelity required for a given problem. For many problems such as aircraft simulation, it may be sufficient to consider gravity to be a constant, defined as: \cite{Stevens2003}

\begin{equation}
\
g = 9.80665  m  (32.1740 ft)/s^{2}
\
\end{equation}

based upon data from World Geodetic System 1984 (WGS-84) \cite{Geodetic}, where g is understood to be pointing (down) in the local frame of reference.

If it is desirable to model an object's weight on Earth as a function of latitude, one could use the following \cite{Stevens2003}:
\begin{equation}
\
g=g_{45} - \tfrac{1}{2}(g_{\mathrm{poles}}-g_{\mathrm{equator}}) \cos\left(2\, lat\, \frac{\pi}{180}\right)
\
\end{equation}

where
\begin{equation}
\
g_{\mathrm{poles}} = 9.832 m (32.26 ft) / s^{2}
\
\end{equation}

\begin{equation}
\
g_{45} = 9.806 m (32.17 ft) / s^{2}
\
\end{equation}

\begin{equation}
\
g_{\mathrm{equator}} = 9.780 m (32.09 ft) / s^{2}
\
\end{equation}

lat = latitude, between -90$^{\circ}$ and 90$^{\circ}$. \\

Both these models take into account the centrifugal relief that is produced by the rotation of the Earth, and neither accounts for changes in gravity with changes in altitude. It is worth noting that for the mass attraction effect by itself, the gravitational acceleration at the equator is about 0.18\% less than that at the poles due to being located farther from the mass center. When the rotational component is included (as above), the gravity at the equator is about 0.53\% less than that at the poles, with gravity at the poles being unaffected by the rotation. So the rotational component change due to latitude (0.35\%) is about twice as significant as the mass attraction change due to latitude (0.18\%), but both reduce strength of gravity at the equator as compared to gravity at the poles.\\

Note that for satellites, orbits are decoupled from the rotation of the Earth so the orbital period is not necessarily one day, but also that errors can accumulate over multiple orbits so that accuracy is important. For such problems, the rotation of the Earth would be immaterial unless variations with longitude are modeled. Also, the variation in gravity with altitude becomes important, especially for highly elliptical orbits.\\

The \textit{Earth Gravitational Model 1996} (EGM96) contains 130,676 coefficients that refine the model of the Earth's gravitational field \cite{Stevens2003}. The most significant correction term is about two orders of magnitude more significant than the next largest term \cite{Stevens2003}. That coefficient is referred to as the \textit{J}$_2$ term, and accounts for the flattening of the poles, or the oblateness, of the Earth. (A shape elongated on its axis-of-symmetry, like an American football, would be called prolate.) A gravitational potential function can be written for the change in potential energy for a unit mass that is brought from infinity into proximity to the Earth. Taking partial derivatives of that function with respect to a coordinate system will then resolve the directional components of the gravitational acceleration vector, as a function of location. The component due to the Earth's rotation can then be included, if appropriate, based on a sidereal day relative to the stars ($\approx$366.24 days/year) rather than on a solar day ($\approx$365.24 days/year). That component is perpendicular to the axis of rotation rather than to the surface of the Earth.\\

A similar model adjusted for the geometry and gravitational field for Mars can be found in publication NASA SP-8010. \cite{Noll1974}\\

The barycentric gravitational acceleration at a point in space is given by:
\begin{equation}
\
\mathbf{g}=-{G M \over r^2}\mathbf{\hat{r}}
\
\end{equation}

where:

\textit{M} is the mass of the attracting object, $\vec{r}$ is the unit vector from center-of-mass of the attracting object to the center-of-mass of the object being accelerated, \textit{r} is the distance between the two objects, and \textit{G} is the gravitational constant.\\

When this calculation is done for objects on the surface of the Earth, or aircraft that rotate with the Earth, one has to account that the Earth is rotating and the centrifugal acceleration has to be subtracted from this. For example, the equation above gives the acceleration at 9.820 m/s$^{2}$, when \textit{GM} = 3.986$\times$10$^{14}$ m$^{2}$/s$^{2}$, and \textit{R} = 6.371$\times$10$^{6}$ m. The centripetal radius is \textit{r = R} cos(latitude), and the centripetal time unit is approximately (day / 2π), reduces this, for \textit{r} = 5$\times$10$^{6}$ metres, to 9.79379 m/s$^{2}$, which is closer to the observed value.\\

\section{Speed light}

The 1889 definition of the metre, based on the international prototype of platinum-iridium, was replaced by the 11th CGPM (1960) using a definition based on the wavelength of krypton 86 radiation. This change was adopted in order to improve the accuracy with which the definition of the metre could be realized, the realization being achieved using an interferometer with a travelling microscope to measure the optical path difference as the fringes were counted. In turn, this was replaced in 1983 by the 17th CGPM (1983, Resolution 1) that specified the current definition, as follows:\\

The metre is the length of the path travelled by light in vacuum during a time interval of 1/299,792,458 of a second.\\

It follows that the speed of light in vacuum is exactly 299,792,458 metres per second,

\begin{equation}
\
c_{0} = 299,792,458 m/s.
\
\end{equation}

The original international prototype of the metre, which was sanctioned by the 1st CGPM in 1889, is still kept at the BIPM under conditions specified in 1889. \cite{(BIPM)2006-2014}

The speed of light in vacuum, commonly denoted \textit{c}, is a universal physical constant important in many areas of physics. Its value is exactly 299,792,458 metres per second

\begin{equation}
\
c \approx 3.00 \times 10^{8} m/s,
\
\end{equation}

as the length of the metre is defined from this constant and the international standard for time \cite{Penrose2004}. According to special relativity, c is the maximum speed at which all matter and information in the universe can travel. It is the speed at which all massless particles and changes of the associated fields (including electromagnetic radiation such as light and gravitational waves) travel in vacuum. Such particles and waves travel at c regardless of the motion of the source or the inertial reference frame of the observer. In the theory of relativity, \textit{c} interrelates space and time, and also appears in the famous equation of mass–energy equivalence  \textit{E} = \textit{mc}$^{2}$ \cite{Einstein, Uzan2008}\\

The speed at which light propagates through transparent materials, such as glass or air, is less than \textit{c}; similarly, the speed of radio waves in wire cables is slower than \textit{c}. The ratio between \textit{c} and the speed \textit{v} at which light travels in a material is called the refractive index \textit{n} of the material

\begin{equation}
\
n = c / v.
\
\end{equation}

For example, for visible light the refractive index of glass is typically around 1.5, meaning that light in glass travels at $c / 1.5$ $\approx$ 200,000 km/s; the refractive index of air for visible light is about 1.0003, so the speed of light in air is about 299,700 km/s (about 90 km/s slower than \textit{c}).\\

For many practical purposes, light and other electromagnetic waves will appear to propagate instantaneously, but for long distances and very sensitive measurements, their finite speed has noticeable effects. In communicating with distant space probes, it can take minutes to hours for a message to get from Earth to the spacecraft, or vice versa. The light seen from stars left them many years ago, allowing the study of the history of the universe by looking at distant objects. The finite speed of light also limits the theoretical maximum speed of computers, since information must be sent within the computer from chip to chip. The speed of light can be used with time of flight measurements to measure large distances to high precision.\\

Ole R$\phi$mer first demonstrated in 1676 that light travels at a finite speed (as opposed to instantaneously) by studying the apparent motion of Jupiter's moon Io. In 1865, James Clerk Maxwell proposed that light was an electromagnetic wave, and therefore travelled at the speed c appearing in his theory of electromagnetism \cite{Gibbs1997}. In 1905, Albert Einstein postulated that the speed of light with respect to any inertial frame is independent of the motion of the light source \cite{Stachel2002}, and explored the consequences of that postulate by deriving the special theory of relativity and showing that the parameter \textit{c} had relevance outside of the context of light and electromagnetism.\\

After centuries of increasingly precise measurements, in 1975 the speed of light was known to be 299,792,458 m/s with a measurement uncertainty of 4 parts per billion. In 1983, the metre was redefined in the International System of Units (SI) as the distance travelled by light in vacuum in 1/299,792,458 of a second. As a result, the numerical value of \textit{c} in metres per second is now fixed exactly by the definition of the metre. \cite{Resnick, (BIPM)2006-2014, speedlight} \\

The speed of electromagnetic wave (c) is:

\begin{equation}
\
c = \frac{1}{\sqrt{\mu_0 \varepsilon_0}} 
\
\end{equation}

In 1865 Maxwell wrote:\\

``This speed is so close to the speed of light that it seems we have strong reason to conclude that light itself (including radiant heat, and other radiations if any) is an electromagnetic disturbance in the form of waves propagated through the electromagnetic field according to electromagnetic laws." \cite{Resnick, speedlight, Maxwell1878}

\section{Mass-energy equivalence}

In physics, mass–energy equivalence is the concept that the mass of an object or system is a measure of its energy content. For instance, adding 25 kWh (90 MJ) of any form of energy to any object increases its mass by 1 $\mu$g (and, accordingly, its inertia and weight) even though no matter has been added.

A physical system has a property called energy and a corresponding property called mass; the two properties are equivalent in that they are always both present in the same (i.e. constant) proportion to one another. Mass–energy equivalence arose originally from special relativity as a paradox described by Henri Poincar\'e \cite{Poincare1900}. The equivalence of energy \textit{E} and mass \textit{m} is reliant on the speed of light \textit{c} and is described by the famous equation:
\begin{equation}
\
E = mc^2
\
\end{equation}

Thus, this mass–energy relation states that the universal proportionality factor between equivalent amounts of energy and mass is equal to the speed of light squared. This also serves to convert units of mass to units of energy, no matter what system of measurement units used.\\

If a body is stationary, it still has some internal or intrinsic energy, called its rest energy. Rest mass and rest energy are equivalent and remain proportional to one another. When the body is in motion (relative to an observer), its total energy is greater than its rest energy. The rest mass (or rest energy) remains an important quantity in this case because it remains the same regardless of this motion, even for the extreme speeds or gravity considered in special and general relativity; thus it is also called the invariant mass.\\

On the one hand, the equation \textit{E} = \textit{mc}$^{2}$ can be applied to rest mass (\textit{m} or \textit{m}$_{0}$) and rest energy (\textit{E}$_{0}$) to show their proportionality as \textit{E}$_{0}$ = \textit{m}$_{0}$\textit{c}$^{2}$  \cite{B.Okun1969}.\\

On the other hand, it can also be applied to the total energy (\textit{E}$_{tot}$ or simply \textit{E}) and total mass of a moving body. The total mass is also called the relativistic mass mrel. The total energy and total mass are related by \textit{E} = \textit{m}$_{rel}$\textit{c}$^{2}$. \cite{PaulAllenTipler2003}\\

Thus, the mass–energy relation \textit{E} = \textit{mc}$^{2}$ can be used to relate the rest energy to the rest mass, or to relate the total energy to the total mass. To instead relate the total energy or mass to the rest energy or mass, a generalization of the mass–energy relation is required: the energy–momentum relation.\\

\textit{E} = \textit{mc}$^{2}$ has frequently been invoked as an explanation for the origin of energy in nuclear processes specifically, but such processes can be understood as converting nuclear potential energy in a manner precisely analogous to the way that chemical processes convert electrical potential energy. The more common association of mass–energy equivalence with nuclear processes derives from the fact that the large amounts of energy released in such reactions may exhibit enough mass that the mass loss (which is called the mass defect) may be measured, when the released energy (and its mass) have been removed from the system; while the energy released in chemical processes is smaller by roughly six orders of magnitude, and so the resulting mass defect is much more difficult to measure. For example, the loss of mass to an atom and a neutron, as a result of the capture of the neutron and the production of a gamma ray, has been used to test mass–energy equivalence to high precision, as the energy of the gamma ray may be compared with the mass defect after capture. In 2005, these were found to agree to 0.0004\%, the most precise test of the equivalence of mass and energy to date. This test was performed in the World Year of Physics 2005, a centennial celebration of Albert Einstein's achievements in 1905. \cite{Nature2005}\\

Einstein was not the first to propose a mass–energy relationship (see the History section). However, Einstein was the first scientist to propose the \textit{E} = \textit{mc}$^{2}$ equation and the first to interpret mass–energy equivalence as a fundamental principle that follows from the relativistic symmetries of space and time. \cite{Mass}

\subsection{Albert Einstein}

Albert Einstein (14 March 1879 - 18 April 1955) was a German-born theoretical physicist. He developed the general theory of relativity, one of the two pillars of modern physics (alongside quantum mechanics) \cite{Whittaker1955, Yang2010}. Einstein's work is also known for its influence on the philosophy of science \cite{Howard2004-2014, Howard2005}. Einstein is best known in popular culture for his mass–energy equivalence equation \textit{E} = \textit{mc}$^{2}$ (which has been dubbed ``the world's most famous equation") \cite{Bodanis2000}. He received the 1921 Nobel Prize in Physics for his ``services to theoretical physics", in particular his discovery of the law of the photoelectric effect, a pivotal step in the evolution of quantum theory. \cite{Foundation1921}\\

\begin{figure}
	\begin{center}
		\includegraphics[scale=0.3]{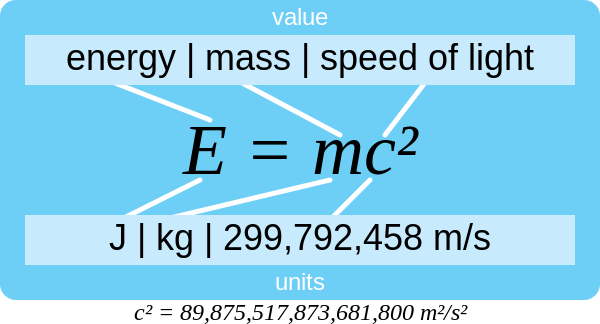}
		\caption{\small{Photography Albert Einstein and Equation    \textit{E} = \textit{mc}$^{2}$ explanation. \cite{Mass, Einstein}}}\label{fig:sweeps}
	\end{center}
\end{figure}

Near the beginning of his career, Einstein thought that Newtonian mechanics was no longer enough to reconcile the laws of classical mechanics with the laws of the electromagnetic field. This led to the development of his special theory of relativity. He realized, however, that the principle of relativity could also be extended to gravitational fields, and with his subsequent theory of gravitation in 1916, he published a paper on general relativity. He continued to deal with problems of statistical mechanics and quantum theory, which led to his explanations of particle theory and the motion of molecules. He also investigated the thermal properties of light which laid the foundation of the photon theory of light. In 1917, Einstein applied the general theory of relativity to model the large-scale structure of the universe. \cite{Nobel2011}\\

Einstein published more than 300 scientific papers along with over 150 non-scientific works \cite{Nobel2011}. Einstein's intellectual achievements and originality have made the word ``Einstein" synonymous with ``genius". \cite{Einstein}

\subsection{Conservation of mass and energy}

Mass and energy can be seen as two names (and two measurement units) for the same underlying, conserved physical quantity. Einstein was unequivocally against the traditional idea of conservation of mass. He had concluded that mass and energy were essentially one and the same; inertial mass is simply latent energy. He made his position known publicly time and again, \textit{Eugene Hecht} \cite{Hecht2009}. Thus, the laws of conservation of energy and conservation of (total) mass are equivalent and both hold true ``There followed also the principle of the equivalence of mass and energy, with the laws of conservation of mass and energy becoming one and the same", \textit{Albert Einstein} \cite{Einstein1940}. Einstein elaborated in a 1946 essay that ``the principle of the conservation of mass", proved inadequate in the face of the special theory of relativity. It was therefore merged with the energy [conservation] principle-just as, about 60 years before, the principle of the conservation of mechanical energy had been combined with the principle of the conservation of heat [thermal energy]. ``We might say that the principle of the conservation of energy, having previously swallowed up that of the conservation of heat, now proceeded to swallow that of the conservation of mass—and holds the field alone." \cite{Einstein1940}

If the conservation of mass law is interpreted as conservation of rest mass, it does not hold true in special relativity. The rest energy (equivalently, rest mass) of a particle can be converted, not ``to energy" (it already is energy (mass)), but rather to other forms of energy (mass) which require motion, such as kinetic energy, thermal energy, or radiant energy; similarly, kinetic or radiant energy can be converted to other kinds of particles which have rest energy (rest mass). In the transformation process, neither the total amount of mass nor the total amount of energy changes, since both are properties which are connected to each other via a simple constant \cite{Stanford, Taylor1992}. This view requires that if either energy or (total) mass disappears from a system, it will always be found that both have simply moved off to another place, where they may both be measured as an increase of both energy and mass corresponding to the loss in the first system.

\subsection{Application to nuclear physics}

Max Planck pointed out that the mass-energy equivalence equation implied that bound systems would have a mass less than the sum of their constituents, once the binding energy had been allowed to escape. However, Planck was thinking about chemical reactions, where the binding energy is too small to measure. Einstein suggested that radioactive materials such as radium would provide a test of the theory, but even though a large amount of energy is released per atom in radium, due to the half-life of the substance (1602 years), only a small fraction of radium atoms decay over an experimentally measurable period of time.

Once the nucleus was discovered, experimenters realized that the very high binding energies of the atomic nuclei should allow calculation of their binding energies, simply from mass differences. But it was not until the discovery of the neutron in 1932, and the measurement of the neutron mass, that this calculation could actually be performed (see nuclear binding energy for example calculation). A little while later, the first transmutation reactions such as \cite{Cockroft} the \textit{Cockcroft-Walton} experiment: 

\begin{equation}
\
{}_{4}^{7} {Li}+\, {}_{1}^{1} H\, \, \to \, \, 2\, {}_{2}^{4} {He}\, 
+Energy
\
\end{equation}
 
verified Einstein's equation to an accuracy of $\pm$0.5\%. In 2005, \textit{Rainville et al} \cite{Nature2005}. published a direct test of the energy-equivalence of mass lost in the binding energy of a neutron to atoms of particular isotopes of silicon and sulfur, by comparing the mass lost to the energy of the emitted gamma ray associated with the neutron capture. The binding mass-loss agreed with the gamma ray energy to a precision of $\pm$0.00004\%, the most accurate test of\\ \textit{E} = \textit{mc}$^{2}$ to date. \cite{Nature2005}

The mass–energy equivalence equation was used in the understanding of nuclear fission reactions, and implies the great amount of energy that can be released by a nuclear fission chain reaction, used in both nuclear weapons and nuclear power. By measuring the mass of different atomic nuclei and subtracting from that number the total mass of the protons and neutrons as they would weigh separately, one gets the exact binding energy available in an atomic nucleus. This is used to calculate the energy released in any nuclear reaction, as the difference in the total mass of the nuclei that enter and exit the reaction. \cite{Mass}

\subsection{Mass in special relativity}

Mass in special relativity incorporates the general understandings from the concept of mass–energy equivalence. Added to this concept is an additional complication resulting from the fact that mass is defined in two different ways in special relativity: one way defines mass (\textit{rest mass} or \textit{invariant mass}) as an invariant quantity which is the same for all observers in all reference frames; in the other definition, the measure of mass (\textit{relativistic mass}) is dependent on the velocity of the observer.\\

The term mass in special relativity usually refers to the rest mass of the object, which is the Newtonian mass as measured by an observer moving along with the object. The invariant mass is another name for the rest mass of single particles. The more general invariant mass (calculated with a more complicated equation) loosely corresponds to the \textit{rest mass} of a \textit{system}. Thus, invariant mass is a natural unit of mass used for systems which are being viewed from their center of momentum frame, as when any closed system (for example a bottle of hot gas) is weighed, which requires that the measurement be taken in the center of momentum frame where the system has no net momentum. Under such circumstances the invariant mass is equal to the relativistic mass, which is the total energy of the system divided by \textit{c} (the speed of light) squared.\\

The concept of invariant mass does not require bound systems of particles, however. As such, it may also be applied to systems of unbound particles in high-speed relative motion. Because of this, it is often employed in particle physics for systems which consist of widely separated high-energy particles. If such systems were derived from a single particle, then the calculation of the invariant mass of such systems, which is a never-changing quantity, will provide the rest mass of the parent particle (because it is conserved over time).\\

It is often convenient in calculation that the invariant mass of a system is the total energy of the system (divided by \textit{c}$^{2}$) in the center of momentum frame (where, by definition, the momentum of the system is zero). However, since the invariant mass of any system is also the same quantity in all inertial frames, it is a quantity often calculated from the total energy in the center of momentum frame, then used to calculate system energies and momenta in other frames where the momenta are not zero, and the system total energy will necessarily be a different quantity than in the  center of momentum frame. As with energy and momentum, the invariant mass of a system cannot be destroyed or changed, and it is thus conserved, so long as the system is closed to all influences (The technical term is isolated system meaning that an idealized boundary is drawn around the system, and no mass/energy is allowed across it).\\

\textbf{Invariant mass}\\

The invariant mass is the ratio of four-momentum to four-velocity: \cite{McGlinn2004}

\begin{equation}
\
p^\mu = m v^\mu\
\
\end{equation}

and is also the ratio of four-acceleration to four-force when the rest mass is constant. The four-dimensional form of Newton's second law is:

\begin{equation}
\
F^\mu = m A^\mu.\!
\
\end{equation}

\textbf{The relativistic energy-momentum equation}\\

The relativistic expressions for \textit{E} and \textit{p} obey the relativistic energy–momentum relation: \cite{Taylor1992a}

\begin{equation}
E^2 - (pc)^2 = (mc^2)^2 \,\!
\end{equation}

where the m is the \textit{rest mass}, or the invariant mass for systems, and \textit{E} is the total energy.

The equation is also valid for photons, which have \textit{m} = \textit{0}:

\begin{equation}
E^2 - (pc)^2 = 0 \,\!
\end{equation}

and therefore

\begin{equation}
E = pc \,\!
\end{equation}

A photon's momentum is a function of its energy, but it is not proportional to the velocity, which is always \textit{c}.
For an object at rest, the momentum \textit{p} is zero, therefore

\begin{equation}
E_0 = mc^2 \,\! 	
\end{equation}
	
\textit{true only for particles or systems with momentum = 0}.
The rest mass is only proportional to the total energy in the rest frame of the object. When the object is moving, the total energy is given by

\begin{equation}
E = \sqrt{ (mc^2)^2 + (pc)^2 } \,\!
\end{equation}

\begin{figure}
	\begin{center}
		\includegraphics[scale=0.66]{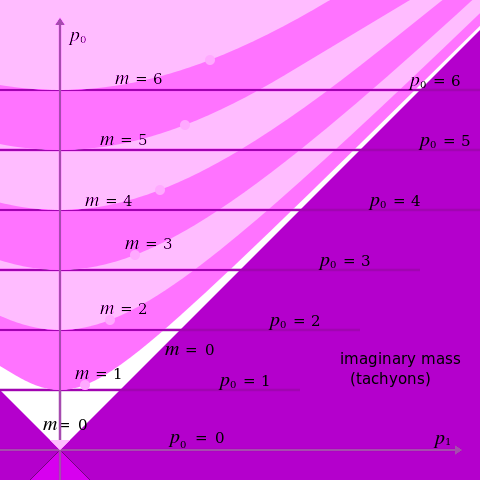}
		\caption{\small{Dependency between the rest mass and \textit{E}, given in 4-momentum (\textit{p}$_{0}$,\textit{p}$_{1}$) coordinates; \textit{p}$_0$\textit{c} = \textit{E} \cite{relativity}.}}\label{fig:sweeps}
	\end{center}
\end{figure}

To find the form of the momentum and energy as a function of velocity, it can be noted that the four-velocity, which is proportional to (\textit{c}, $\vec{v}$), is the only four-vector associated with the particle's motion, so that if there is a conserved four-momentum (\textit{E}, $\vec{pc}$), it must be proportional to this vector. This allows expressing the ratio of energy to momentum as

\begin{equation}
pc=E {v \over c}
\end{equation}

resulting in a relation between E and v:

\begin{equation}
E^2 = (mc^2)^2 + E^2 {v^2\over c^2},
\end{equation}

This results in

\begin{equation}
E = {mc^2 \over \sqrt{1-\displaystyle{v^2\over c^2}}}
\end{equation}

and

\begin{equation}
p = {mv\over \sqrt{1-\displaystyle{v^2\over c^2}}}.
\end{equation}

these expressions can be written as

\begin{equation}
E_0= mc^2 \,,
\end{equation}

\begin{equation}
E=\gamma mc^2 \,,
\end{equation}

and

\begin{equation}
p = mv \gamma \,.
\end{equation}

When working in units where \textit{c} = \textit{1}, known as the natural unit system, all relativistic equations simplify. In particular, all three quantities \textit{E}, \textit{p}, \textit{m} have the same dimension:

\begin{equation}
m^2 = E^2 - p^2 \,\!.
\end{equation}

The equation is often written this way because the difference \textit{E}$^2$ $-$ \textit{p}$^2$ is the relativistic length of the energy momentum four-vector, a length which is associated with rest mass or invariant mass in systems. If \textit{m} $>$ \textit{0}, then there is the rest frame, where \textit{p} = \textit{0}, this equation states that \textit{E} = \textit{m}, revealing once more that invariant mass is the same as the energy in the rest frame.

\section{Methods and calculations}

By Einstein's equation \cite{Einstein, Einstein1940}, Equation (28), matter is converted into energy and vice versa.\\

Whereas a mass of 10 mg and the speed of light, c = 299,792,458 m/s, calculating \textit{E}:

\[
E=mc^{2}
\]

\[
E = 10 \times {10}^{-6}\, kg\, .\, \left( 299,792,458\, m/s \right)^{2}
\]
\begin{equation}
\
E \approx 
8.98755 \times {10}^{11}\, J
\
\end{equation}

Therefore, the energy of 10 mg de matter is
\begin{center}
	 \textit{E} $\approx 8.98755 \times {10}^{11}\, J$.
\end{center}

\subsection{Scaling the size of the mountain that can lift}

\subsubsection{Massive Iron Mountain}

Consider a mountain in dimensional cone-shaped radius of 100m and 100m in height built in massive iron. We know that the iron density is 7.874 g/cm$^{3}$, the volume of a right cone \cite{US2003, MathWorld1999-2015} is given by

\begin{equation}
	\
	V = \frac{1}{3}\pi R^{2} H
	\
\end{equation}

to \textit{R} = \textit{H} have

\begin{equation}
\
V = \frac{1}{3}\pi R^{3},
\
\end{equation}

absolute density that is equal to the ratio between mass and volume of the body, 

\begin{equation}
\
d = \frac{m}{V},
\
\end{equation}

the acceleration of gravity is \textit{g} = 9,80665 m/s$^{2}$ and the gravitational potential energy surface Equation (7),

\[
U = U_{0} + mgy, 
\]

to \textit{U}$_{0}$ = \textit{0}, 

\begin{equation}
\
U = + mgy
\end{equation}

so we can calculate how high can lift one mountain massive iron with the energy of a 10 mg of matter.\\

making the equation (28) equals (49), and by replacing the mass (\textit{m} = \textit{d} .\textit{V}) where V given by equation (47), therefore

\[
E = mc^{2} = U = + mgy \Rightarrow E = d\frac{1}{3}\pi R^{3}g y \Rightarrow
\]

\begin{equation}
\
y = \frac{3E}{d\pi R^{3} g} 
\
\end{equation}

that is the height y which can lift a mountain in the form of massive iron cone, with the mass energy of 10 mg of matter.\\

For m = 10 mg, one has  \textit{E} $\approx 8.98755 \times {10}^{11}\, J$, therefore:

\begin{equation}
\
y = \frac{3 \ldotp (8.98755 \times {10}^{11}\, J)}{7,854 kg/m^{3} . (3.14159) \ldotp (100m)^{3} \ldotp 9.80665 m/s^{2}} 
\
\end{equation}

\begin{equation}
\
\Rightarrow y \cong 11.142974 m  \cong 1.11 \times 10^{1} m
\
\end{equation}

\subsubsection{Rising Mount Everest}

As the Figures (1), (2) and (6), the shape of Mount Everest is an equilateral triangular pyramid whose side of the triangular base is about 5,200 m, and the height from the base camp is 3,548 m to its top.\\

Similarly we calculated how high we can build a massive iron mountain in the previous item, we will calculate how high we raise the mount everest, watching him from his Figure (6) camping base, only with the energy of 10 mg of matter.\\

Let us consider the formation of basalt mountain, whose density is 2.7 g/cm$^{3}$, the acceleration of gravity starting from the surface \textit{g} = 9.80665 m/s$^{2}$ and the volume of a pyramid  \cite{US2003, MathWorld1999-2015} is given as:\\

\begin{equation}
\
V = \frac{1}{3}(Base \ area) \times  H
\
\end{equation}

Then we calculate first the area of the base of the Mount Everest, seen here in the shape of a pyramid with equilateral triangular base. As the area of a triangle  \cite{US2003, MathWorld1999-2015} is given by:

\begin{equation}
\
A_{(\Delta)} = \frac{Base \times Height}{2} 
\
\end{equation}

\begin{equation}
\
A_{(\Delta)} = \frac{5,200 m \times 4,503.332 m }{2} \approxeq  11,708,663.2 m^{2}
\
\end{equation}
area triangular base of the Mount Everest.\\

The volume will be given by:\\

\[
V = \frac{1}{3}(Base \ area) \times  H 
\]
\begin{equation}
\
= \frac{1}{3}(11,708,663.2 m^{2}) \times  3,548 m
\end{equation}

\begin{equation}
\Rightarrow V \cong 13,847,445,677.867 m^{3} \cong 1.385 \times 10^{10} m^{3}
\end{equation}

\begin{figure}
	\begin{center}
		\includegraphics[scale=0.47]{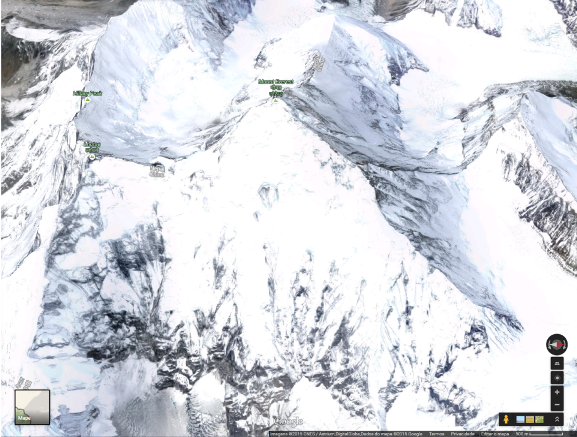}\\
		\includegraphics[scale=0.37]{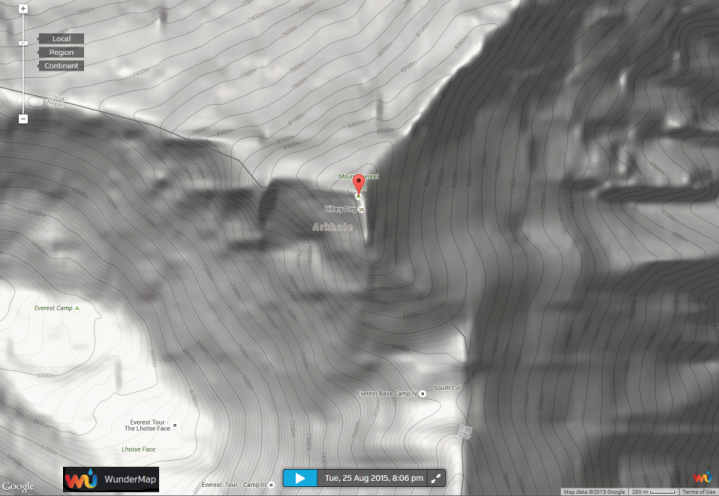}
		\caption{\small{Mount Everest, 27$^{\circ}$59'20.9"N 86$^{\circ}$55'31.3"E, Google Map \cite{googlemap, WunderMap}.}}\label{fig:sweeps}
	\end{center}
\end{figure}

Now calculating how high \textit{y} we raise the Mount Everest, we have:\

\begin{equation}
\
E = mc^{2} = U = + mgy \Rightarrow E = dVgy 
\
\end{equation}

\begin{equation}
\
\Rightarrow y = \frac{E}{dVg}
\
\end{equation}

for m = 10 mg, one has  \textit{E} $\approx 8.98755 \times {10}^{11}\, J$, therefore:

\[
y = \frac{E}{dVg}
\]

\begin{equation}
\
y = \frac{8.98755 \times {10}^{11}\, J}{2,700 kg/m^{3} . (13,847,445,677.867 m^{3}) \ldotp 9.80665 m/s^{2}} 
\
\end{equation}

\begin{equation}
\
\Rightarrow y \cong 0.00245124776 mm \cong 2.45  \times 10^{0}mm
\
\end{equation}

\section{Conclusions} 
Energy is energy, no matter how it is. As conserved energy can be used in forms diversar.
The calculations showed that despite a small mass, 10 mg, when converted fully energy by the equation \textit{E} = \textit{mc}$^{2}$, it generates a lot of energy. This may raise a massive iron mountain, cone-shaped, 100 m  height, to 11.14 meters above the ground. Going further, also can be with the same energy, lift Mount Everest to 2.45 millimeters above the ground.

\bibliographystyle{unsrt}
\bibliography{journals}

\end{document}